\documentclass[runningheads]{llncs}
\usepackage[T1]{fontenc}
\usepackage{graphicx}
\usepackage[english]{babel}\addto\captionsenglish  
{}  
\usepackage{blindtext}
\usepackage[sorting=none]{biblatex}
\usepackage{tabularx}
\usepackage[utf8]{inputenc}
\usepackage{hyperref}
\usepackage[caption=false]{subfig}
\usepackage{graphicx}
\usepackage{mathtools}
\usepackage{pifont}
\usepackage{comment}
\usepackage{algorithm}
\usepackage{mathtools}
\usepackage{authblk}
\usepackage{amsmath}
\usepackage{algorithm}
\usepackage{algorithmic}
\usepackage{tabulary} 
\usepackage{booktabs} 
\usepackage{multirow} 
\usepackage{array} 
\usepackage{ragged2e}
\begin{document}
\title{Enhancing Workflow Security in Multi-Cloud Environments through Monitoring and Adaptation upon Cloud Service and Network Security Violations}
\titlerunning{Enhancing Workflow Security in Multi-Cloud Environments}
%
%

\author{Nafiseh Soveizi\inst{1}\orcidID{0000-0003-2111-734X} \and
Dimka Karastoyanova\inst{1}\orcidID{0000-0002-8827-2590}}
\authorrunning{N.Soveizi and D.Karastoyanova}
%
\institute{Information Systems Group, University of Groningen, Groningen, The Netherlands
\email{\{n.soveizi,d.karastoyanova\}@rug.nl}}
\maketitle            
\begin{abstract}
Cloud computing has emerged as a crucial solution for handling data- and compute-intensive workflows, offering scalability to address dynamic demands. However, ensuring the secure execution of workflows in the untrusted multi-cloud environment poses significant challenges, given the sensitive nature of the involved data and tasks. The lack of comprehensive approaches for detecting attacks during workflow execution, coupled with inadequate measures for reacting to security and privacy breaches has been identified in the literature. To close this gap, in this work, we propose an approach that focuses on monitoring cloud services and networks to detect security violations during workflow executions. Upon detection, our approach selects the optimal adaptation action to minimize the impact on the workflow.
To mitigate the uncertain cost associated with such adaptations and their potential impact on other tasks in the workflow, we employ adaptive learning to determine the most suitable adaptation action. 
Our approach is evaluated based on the performance of the detection procedure and the impact of the selected adaptations on the workflows.


\end{abstract}
\section{Introduction}
\vspace*{-.2cm}
Cloud computing has emerged as a vital solution for organizations dealing with data- and compute-intensive workflows, offering unparalleled scalability and flexibility to meet dynamic demands. By providing a platform for outsourcing workflow execution and storage, the cloud has revolutionized the way organizations operate and cooperate. However, despite all the advantages of cloud-based workflows, cloud security is a major area of concern \cite{Varshney2019, Maguluri2012}, limiting its adoption for workflows involving sensitive data and tasks. 

The distributed nature of workflows allows for dynamic binding to cloud services, which can lead to increased security risks and vulnerability to malicious attacks, as these services may encounter security issues that were unknown during the modelling or even during the binding phase. Additional security-related challenges are introduced by the transmission of sensitive data among cloud components, such as Data Centers (DCs), over potentially untrusted network channels. Therefore, it is crucial to closely monitor the behavior of cloud services and network infrastructure in order to detect and react to any potential violations. Towards this goal, in this paper, we propose an approach that focuses on monitoring, detecting, and reacting to security violations during workflow execution, focusing on cloud services and network violations through the analysis of network traffic data and log files received from cloud providers.


The subsequent step of reacting to the detected security violations boils down to selecting the appropriate adaptation action to minimize the impact of these detected violations. This task is complex due to the presence of \textit{uncertain overhead costs associated with each adaptation action}, which cannot be accurately determined during the workflow modeling, scheduling, or even when reacting to detected violations. These uncertainties vary across different workflow types. For example, analyzing past instances of reworking tasks in a particular workflow reveals that certain task types tend to have more uncertain delays when reworked compared to others. Similarly, examining the consequences of skipping tasks in previous workflow instances highlights the potential negative effects on other tasks, even leading to failures. 

These uncertain costs within the workflow are closely connected to several key factors: 1) \textit{The current state of the workflow} has a significant influence on the potential risks and uncertain costs. For instance, if the workflow is already experiencing delays due to the dynamic nature of cloud performance \cite{chen2015towards} or has encountered multiple violations, the costs of reworking or resequencing tasks can be higher due to the requirement of additional resources or potential disruptions to ongoing tasks. 2) \textit{The previously accrued violations and their respective adaptations} play a crucial role in determining the uncertain costs. Each violation and its adaptation can have a cascading effect on the entire workflow, impacting subsequent tasks and introducing further uncertainties. Considerations such as dependencies and compatibility issues need to be taken into account when deciding on the next adaptation action. 3)  \textit{Workflow complexity} which includes the number of possible tasks, branching paths, and potential variants \cite{nolle2018analyzing}, also contributes to the uncertain costs. Tasks often involve conditional instructions that lead to multiple program branches and loops. The variations in these branches or loops result in diverse task computations, varying execution times, and different outcomes based on different data inputs \cite{chen2016uncertainty}. Hence, the larger the number of possible tasks, branching paths, and potential variants, the greater the uncertainties associated with different inputs, making the estimation of uncertain costs more challenging.

Therefore, there is a need for a method that effectively addresses these uncertainties associated with each adaptation action, particularly when such uncertainties cannot be determined at the time of adaptation action selection. To address this need, our approach is based on learning from past adaptations of workflows to predict the most suitable adaptation action. We consider the uncertain cost of each action and its potential impact on other tasks, taking a holistic perspective that considers the entire workflow at runtime. This approach mitigates risks, supports decision-making, and enhances the system's ability to proactively respond to security violations.

Our approach is based on the \textit{SecFlow} \cite{SecFlow} architecture that enables adaptation on two levels -- tenant level and middleware level -- to ensure a balance between security and efficiency. Our solution separates workflow instances of different tenants, thus meeting their specific functional and non-functional requirements within isolated environments. 
This model incorporates a logically centralized middleware, which facilitates informed decision-making for all tenants and simplifies the cloud infrastructure, thereby hiding complexity from its tenants while minimizing the amount of information possessed by the middleware regarding individual tenants.

The rest of this paper is organized as follows: Section \ref{sec:Related works} provides an overview of the existing monitoring and adaptation mechanisms for security violations in cloud-based workflows. In Section \ref{sec:System Overview}, we present the architecture upon which our proposed method is built. Section \ref{sec:Proposed solution} describes our proposed adaptive approach for monitoring and adapting cloud services and networks to mitigate security violations. Section \ref{sec:Evaluation} presents the evaluation of the proposed approach. Finally, Section \ref{sec:Conclusion} concludes the paper and outlines potential future research directions.

\section{Related works}
\label{sec:Related works}
In this section, we give a brief overview of the existing WfMS featuring monitoring and/or adaptation mechanisms for security violations in multi-cloud environments. We base this overview on a recent systematic review of the state of the art in security and privacy of cloud-based workflows that considers both business and scientific workflows~\cite{soveizi2023security}. 
 \begin{table*}[h!]
\fontsize{8pt}{10pt}\selectfont
\caption{{Comparison of existing cloud-based WfMSs regarding their abilities to monitor and react to security violations.} }
\label{table1}
\def\arraystretch{1}
\ignorespaces 
\vspace*{-.3cm}
\begin{tabulary}{\linewidth}{>{\centering\arraybackslash}p{1.4cm}>{\centering\arraybackslash}p{1.5cm}>{\centering\arraybackslash}p{2cm}>{\centering\arraybackslash}p{2.0cm}>{\centering\arraybackslash}p{2.4cm}>{\centering\arraybackslash}p{2.4cm}}
\hline
\textbf{Paper} & \textbf{Workflow Type} & \textbf{Monitoring Module} & \textbf{Considered Security Objectives} & \textbf{Considered \newline Attacks} &  \textbf{Adaptation \newline Options} \\
\hline
\cite{HosseiniShirvani2020}, 2020 & Business & Cloud-side Monitoring & CIA & VM-based, \newline Network attacks & Static Trust \newline Calculation \\
\hline
\cite{Wang2020b}, 2020 \newline \cite{Wang2021}, 2021 & Scientific & Engine-side Monitoring & I & VM-based \newline attacks & Redundancy \\
\hline
\cite{Wen2020b}, 2020 & Scientific & Cloud-side Monitoring & A & Clouds fail & Rescheduling the uncompleted tasks \\
\hline
\cite{Abazari2019}, 2018 & Scientific & Cloud-side Monitoring & CIA & VM-based \newline attacks & Rescheduling the affected tasks \\
\hline
\cite{Ahmad2021}, 2021 & Scientific & Engine-side Monitoring & A & Hardware and \newline Software faults & Re-work \\
\hline
\cite{Alaei2021b}, 2021 & Scientific & Engine-side Monitoring & A & Unavailability \newline of VMs & Re-work \\
\hline
\end{tabulary}
\vspace*{-.4cm}
\end{table*}

We conducted a comparison of existing research in the field, as summarized in Table \ref{table1}. The majority of these studies primarily focus on scientific workflows, with some utilizing cloud-side monitoring, which raises concerns regarding its full trustworthiness. On the other hand, works that solely rely on engine-side monitoring tend to narrow their focus on task failures \cite{Ahmad2021} or specific types of violations \cite{Wang2020b, Alaei2021b}. In addition, only two papers encompass all three essential security objectives: Confidentiality, Integrity, and Availability (CIA) \cite{HosseiniShirvani2020, Abazari2019}. One significant limitation observed across these works is the absence of a structured solution for adaptation, as they often address only one type of reaction to detected violations. As a result, none of the existing approaches comprehensively tackle all potential attacks that could compromise the CIA of outsourced workflow tasks in multi-cloud environments. Furthermore, the available adaptation actions do not sufficiently mitigate the risks associated with various types of violations. For a more in-depth analysis of these studies, please refer to the original sources or consult the survey in \cite{soveizi2023security}.
 \section{System Overview}
\label{sec:System Overview}
\label{sec:System Overview}
\begin{figure*}[t]
\setlength{\belowcaptionskip}{-12pt}
  \includegraphics[width=\textwidth]{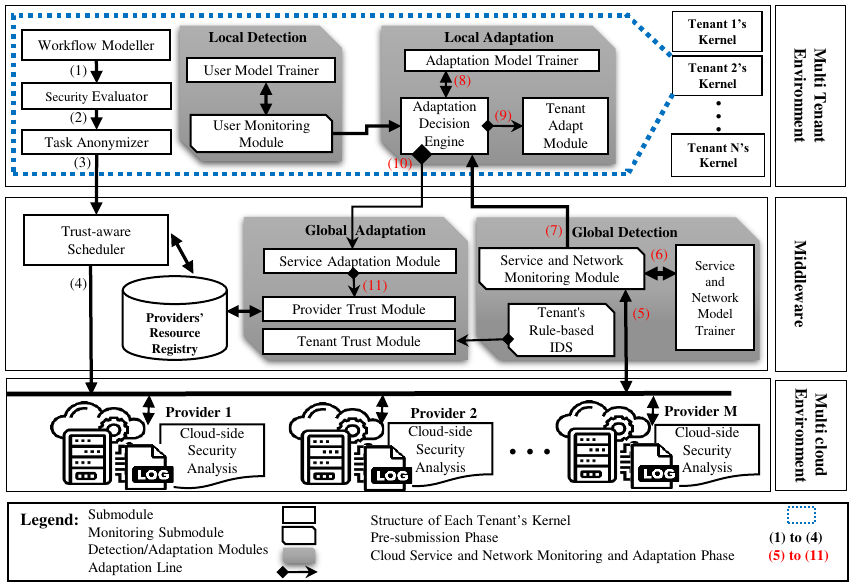}
  \caption{The architecture of SecFlow}
  \label{fig:Architecture}
\end{figure*}

This section presents a brief overview of our security-aware Workflow Management System (WfMS), called \textit{SecFlow} \cite{SecFlow} \footnote{Preprint available at the ArXiv: https://arxiv.org/abs/2307.05137}. It is specifically designed to provide comprehensive protection for workflows throughout their entire lifecycle, safeguarding them against a wide range of security violations and ensuring defense against all potential attackers. Figure \ref{fig:Architecture} depicts the proposed architecture, highlighting key components such as the Tenant's Kernel, the Middleware, and the multi-cloud environment.

In our architecture, we assume that tenants' resources are securely isolated from each other, possibly residing on the same cloud node. The middleware, which acts as a logically centralized component, can be hosted by a trusted third party. To ensure comprehensive monitoring of all potential malicious actors, tenants actively monitor users, while the Middleware oversees Clouds, networks, and tenants, utilizing learned behavioral patterns. In the scope of this paper, we primarily concentrate on identifying and responding to security breaches in cloud services and network infrastructure. In the following, we describe these procedures within the context of the \textit{SecFlow} architecture.


In \textit{SecFlow}, the security-aware management of workflows comprises two main phases: the pre-submission phase (steps 1-4 in Figure \ref{fig:Architecture}) occurring before task submission to cloud environments, and the monitoring and adaptation phase (steps 5-11 in Figure \ref{fig:Architecture}) for detecting and addressing security violations in cloud services and networks.

During the pre-submission phase, tenants utilize the \textbf{\textit{Workflow Modeller}} module to design their workflows under consideration of security requirements (step 1). These workflows are then analyzed using the \textbf{\textit{Security Evaluator}} module to specify potential adaptation actions for each task and incorporate them into the workflow model (step 2). This step is important because certain tasks might have specific eligibility criteria for some actions. For instance, while some tasks may be eligible for skipping, authentication tasks are deemed indispensable for maintaining workflow security. Subsequently, sensitive information is removed from the tasks (step 3) using client-side obfuscation techniques and conflict detection methods. Finally, the \textbf{\textit{Trust-aware Scheduler}} module schedules the workflows, considering tenant requirements and integrating trustworthiness information (step 4). 

In the monitoring and adaptation phase, the \textbf{\textit{Service and Network Monitoring}} module continuously analyzes network traffic and cloud logs to detect malicious activities (step 5), comparing it with expected behavior from the \textbf{\textit{Service and Network Model Trainer}} module (step 6). Detected anomalies trigger alerts to the Adaptation Decision Engine from the corresponding tenant (step 7). The \textbf{\textit{Adaptation Decision Engine}} selects a suitable adaptation option for the detected attack, employing two distinct strategies. The first strategy prioritizes actions that minimize system impact, considering factors like price, time, value, and mitigation score. The second strategy utilizes an adaptive model (step 8) trained by the \textit{Adaptation Model Trainer} module, which takes into account system reactions, current workflow state, and dependencies.

Finally, adaptations are implemented at two levels: tenant- and middleware-level. At the tenant level, the \textbf{\textit{Tenant Adapt Module}} executes tenant-specific adaptation actions (step 9), including Skip, Switch, or Insert \cite{nolle2018binet}. The purpose of this level is to ensure tenant privacy and shield them from the complexities of the underlying cloud infrastructure. At the middleware level, adaptations focus on changes in providers and cloud services, including Reconfiguration, Rework, and Redundancy. The \textbf{\textit{Service Adaptation Module}} re-executes tasks (Rework) or executes them redundantly (Redundancy) to meet evolving tenant requirements (step 10), modifying services within the same provider or exploring alternatives from different providers. Configuration adaptation (Reconfiguration) adjusts the settings of specific cloud services to enhance decision-making. The middleware acts as a central point for receiving feedback, establishing trust factors for providers and services. The \textbf{\textit{Provider Trust Module}} updates trust scores based on detected violations (step 11), updating the trust repository accordingly. These updates improve future scheduling decisions, overall security, and efficiency. Additionally, the module may adjust the service and network model to enhance the monitoring of malicious provider behavior.

Due to the extensive scope of the architecture, this paper does not delve into some of the modules, namely the \textit{Tenant's Rule-based Intrusion Detection System (IDS)} for detecting tenant-originated attacks, the \textit{Tenant Trust Module} for reacting to malicious behavior from tenants, and the \textit{Tenant's Detection module (Local Detection)} for detecting and monitoring malicious user behavior using a pre-trained model. The inclusion of these modules in this paper is meant to provide a complete overview of the entire architecture, however, no further discussion will be dedicated to them.

\section{Monitoring and Adaptation of Security Violations}
\label{sec:Proposed solution}
In this section, we introduce our proposed monitoring and adaptation system, specifically designed to identify security violations occurring during the execution of workflows in both network and cloud services.

To establish a clear understanding of the key concepts involved, we begin this section by introducing important definitions (Section \ref{subsection:Definitions}). These definitions serve as the foundation for subsequent formulas and explanations, ensuring comprehension of the proposed solution. In the subsequent sections, we describe the proposed solution for security violation detection and decision-making regarding the best adaptation action.
\subsection{Definitions}
\label{subsection:Definitions}
\textbf{Definition 1 (Workflow)}: A workflow, denoted as $w$, consists of a set of abstract service tasks that are assigned to concrete cloud services for execution during the deployment phase. It is represented as a tuple ($ST$, $D$, $E_c$, $E_d$) where $ST$ is a set of abstract Service Tasks ($t$), $D$ denotes intermediate Data exchanged between workflow tasks, $E_c$ is a set of control Edges that determine the task execution order based on specified conditions ($Conds$) that must be satisfied ($E_c \subset ST\times ST\times Conds(D)$), $E_d$ is a set of data Edges that specify the flow of data between the tasks $E_d \subset ST\times ST\times D$.

\textbf{Definition 2 (Task)}: A task $t$ in $w$ represents an abstract service task that does not refer to any specific concrete cloud service. It is defined as a tuple ($C$, $I$, $A$, $V$, $AA$), where $C$, $I$, and $A$ represent the Confidentiality, Integrity, and Availability requirements of task $t$, respectively. Additionally, $V$ indicates the Value of the task within the overall workflow, reflecting its contribution to the whole workflow value. This parameter is introduced to assess various adaptation actions for the task. For example, if the value assigned to a task is negligible, skipping it will not significantly impact the final result of the workflow. Moreover, $AA$ indicates a set of feasible Adaptation Actions for task $t$, which can include a combination of actions from both Tenant Adaptation Actions ($TAA$) and Middleware Adaptation Actions ($MAA$). Mathematically, this set can be represented as $AA = \{aa \mid aa \in TAA_{t} \cup MAA_{t}\}$.

\textbf{Definition 3 (Tenant Adaptation Action)}: Tenant Adaptation Actions ($TAA_{t}$) refers to the actions that can be taken at the tenant level to minimize the damage caused by a violation in task $t$. Each of these actions denoted as $Taa_{t}$ ($Taa_{t} \in TAA_{t}$) is defined as a tuple ($P$, $T$, $MI$, $V$) that represents the Price, Time, Mitigation Impact, and Value of the adaptation action for task $t$ in the workflow. These parameters are determined during the workflow modeling phase by considering the specific characteristics and requirements of each action, based on the parameters of task $t$ (refer to Table \ref{table:Adaptation Types}).

\textbf{Definition 4 (Middleware Adaptation Action)}: Middleware Adaptation Actions ($MAA_{t}$) represents the actions that can be employed at the middleware level to minimize the damage caused by a violation in task $t$. Each of these actions denoted as $Maa_{t}$ ($Maa_{t} \in MAA_{t}$) is defined as a tuple ($P$, $T$, $MI$, $V$) that indicates the Price, Time, Mitigation Impact, and Value of the adaptation action based on task $t$ parameters in the workflow. The parameters $P$ and $T$ cannot be predefined during the modeling phase due to the dynamic nature of middleware-level adaptation, which depends on the current workflow state and availability of cloud services. Instead, these parameters are dynamically determined at runtime when an attack is detected, using a backup service ($BackupSrc$) for re-executing the violated tasks (refer to Table \ref{table:Adaptation Types}).

\textbf{Definition 5 (Attack)}: An attack $a$ refers to a security violation that can potentially occur during the execution of a workflow $w$ at task $t$, compromising the security of the task that utilizes service $s$. It is defined as a tuple ($C$, $I$, $A$, $AT$, $MA_{H}$, $MA_{M}$, $MA_{L}$), where $C$, $I$, and $A$ represent the impacts of the attack $a$ on Confidentiality, Integrity, and Availability, respectively. $AT$ denotes the Attack Type of $a$. Additionally, $MA_{H}$, $MA_{M}$, and $MA_{L}$ categorize available Mitigation Actions that can reduce the impact of the attack. These levels correspond to High, Medium, and Low severity, indicating appropriate actions to mitigate the attack's impact with a specific severity (refer to Table \ref{table:Attack-Specification}).

\textbf{Definition 6 (Multi-Cloud Environment)}: A multi-cloud environment consists of a set of cloud services provided by different providers $p_1, \ldots, p_m$. It can be represented as $MC=\bigcup_{p_i \in \{p_1,\ldots,p_m\}} \bigcup_{s \in p_{i_{\textbf{services}}}} (s)$. In this definition, a service $s$ is defined as a tuple ($P$, $T$, $C$, $I$, $A$, $AFR$). Here, $P$ and $T$ represent the Price and average response Time of the service, respectively. The levels of Confidentiality, Integrity, and Availability offered by the service are denoted by $C$, $I$, and $A$ respectively. Additionally, $AFR$ represents the Attack Frequency Rate of service $s$ for different types of attacks. It is calculated as follows $AFR = \bigcup_{AT \in \text{{all AttackTypes}}} AFR(AT)$, where $AFR(AT)$ reflects the likelihood of each attack type $AT$ occurring within the service $s$ based on historical data specific to the service.


\textbf{Definition 7 (Scheduling Plan)}: Scheduling Plan ($SP$) is the process of assigning cloud services to specific tasks within a workflow. It involves the binding of a concrete service $s$ to each abstract service task $t$. It can be expressed as $SP(w) = \bigcup_{t \in w_{ST}, s \in MC} Sched(t \to s)$.
\subsection{Service and Network Monitoring}
The service and network monitoring module plays a crucial role in the real-time detection of service and network attacks using data obtained from various sources, including the real-time service information file and network traffic data. An overview of the monitoring procedure can be found in Algorithm \ref{alg:Service Monitoring Algorithm}.

The algorithm employs the trained attack detection model ($\mathcal{M}_{AttackDetection}$) to analyze real-time data and identify potential attacks (line 2). The attack detection model is trained using Algorithm \ref{alg:TraininAttackDetectiongAlgorithm}, elaborated in Section \hyperref[4.2.1 Attack Detection Trainer]{4.2.1} Upon detection of attack $a_k$ in service task $t_i$, the algorithm employs Algorithm \ref{alg:attack_severity_learning} from Section \hyperref[4.2.2Attack Severity Trainer]{4.2.2} to determine the severity of the attack (line 3).
Furthermore, considering rework and redundancy as the potential options for middleware-level adaptation actions ($MAA_{t_i}$), the algorithm aims to find an appropriate backup service, referred to as $BackupSrc_{t_i}$, for $t_i$ (line 4). It computes the parameters $P$, $T$, $MI$, and $V$ associated with this backup service, taking into consideration the currently available cloud service offerings. 
Following this, the algorithm proceeds to the adaptation decision module of the relevant tenant ($relatedTenant$) to determine the most suitable adaptation action in response to the detected attack (line 5). This process will be further explained in Algorithm \ref{alg:Adaptation Action Selection Algorithm} in Section \ref{Adaptation Decision}.
\vspace*{-.4cm}
\begin{algorithm}
\caption{Service Monitoring Algorithm}
\label{alg:Service Monitoring Algorithm}
\begin{algorithmic}[1]
\REQUIRE $\mathcal{N}$: Real-time Data, $severity$: Attack Severity Model
\ENSURE Monitoring and Detecting Attacks in Real-time Data
\WHILE{system is operational}
\IF{$a_k$ is detected on $\mathcal{N}$ based on $\mathcal{M}_{AttackDetection}$ (Algorithm \ref{alg:TraininAttackDetectiongAlgorithm})} 
\STATE $ l_{a_k} \leftarrow severity_{AT_{a_k}}(a_k)$ (based on Algorithm \ref{alg:attack_severity_learning})
\STATE  $BackupSrc_{t_i} \leftarrow$ findBackupServiceParameters($t_i$)
\STATE $relatedTenant$.AdaptationDecisionEngine($a_k$,$l_{a_k}$,$BackupSrc_{t_i}$)(Algorithm \ref{alg:Adaptation Action Selection Algorithm})
\ENDIF
\ENDWHILE
\end{algorithmic}
\end{algorithm}

\noindent\textbf{4.2.1 Attack Detection Trainer} 
\label{4.2.1 Attack Detection Trainer}

This section focuses on training a robust model to effectively detect attacks using Random Forest and Linear Regression machine learning algorithms. Algorithm \ref{alg:TraininAttackDetectiongAlgorithm} provides an overview of the training process. Two datasets are utilized: Network Traffic Data (NTD) and Cloud Log File (CLF). The NTD dataset contains historical records of network traffic data, exchanged between the middleware and cloud services, enabling the identification of potential network attacks. The CLF dataset consists of the cloud log file, providing resource utilization information such as RAM, CPU, and Bandwidth, for various services, and aiding in the detection of attacks targeting the services.
\vspace*{-.4cm}
\begin{algorithm}
\caption{Attack Detection Trainer Algorithm}
\label{alg:TraininAttackDetectiongAlgorithm}
\begin{algorithmic} [1]
\REQUIRE $\mathcal{DS}$: {NTD , CLF} 
\ENSURE  $\mathcal{M}_{AttackDetection}$: Trained Model for Attack Detection
\FORALL{$\mathcal{D}$ in $\mathcal{DS}$}
\STATE $\mathcal{D}{\text{train}}, \mathcal{D}{\text{test}} \gets$ Split($\mathcal{D}$) into training and testing sets
\STATE $\mathcal{M}_{AttackDetection} \gets$ Train(RandomForest($\mathcal{D}{\text{train}}$), LinearRegression($\mathcal{D}{\text{train}}$)) 
\STATE $Accuracy_{\mathcal{M}_
{AttackDetection}} \gets$ TestAlgorithm($\mathcal{M}_{AttackDetection}, \mathcal{D}{\text{test}}$) 
\ENDFOR
\end{algorithmic}
\end{algorithm}

\noindent\textbf{4.2.2 Attack Severity Trainer}
\label{4.2.2Attack Severity Trainer}

This section introduces an approach to learning the attack severity model for multiple attack types, as depicted in Algorithm \ref{alg:attack_severity_learning}. It employs K-means clustering and chi-square feature selection techniques to assign a severity level to each attack.  By considering the distinctive features of each attack type, the algorithm accurately scores their severity. The Chi-Square Feature Selection is applied to identify the most informative features for each attack type. This ensures that the severity scoring incorporates the specific characteristics and patterns associated with different types of attacks. By considering different features for each attack type, the algorithm significantly enhances the accuracy and granularity of the severity assignment.
\vspace*{-.4cm}
\begin{algorithm}
\caption{Attack Severity Learning Algorithm }
\label{alg:attack_severity_learning}
\begin{algorithmic}[1]
\REQUIRE $\mathcal{DS}$: {NTD , CLF} , $AttackTypes$: List of Attack Types
\ENSURE $severity$: Attack Severity Model for all attack types in the $AttackTypes$
\FORALL{$\mathcal{D}$ in $\mathcal{DS}$}
\FORALL{AttackType $AT$ in $AttackTypes$}
\STATE $\mathcal{DS}_{AT}$ $\leftarrow$ Filter ($\mathcal{DS}$) based on $AT$
\STATE $\mathcal{F'}$ $\leftarrow$ Chi-SquareFeatureSelection($\mathcal{DS}_{AT}$)
\STATE Initialize K-means clustering with $K$ clusters using $\mathcal{DS}_{AT}$ and features $\mathcal{F'}$
\STATE Train K-means on $\mathcal{DS}_{AT}$
\FORALL{clusters $c$ in K-means}
\STATE $severity_{{AT}_\mathcal{D}}[c]$ $\leftarrow$ Calculate the mean attack severity in cluster $c$
\ENDFOR
\ENDFOR
\ENDFOR
\end{algorithmic}
\end{algorithm}
\vspace*{-.4cm}
\vspace*{-.4cm}
\subsection{Adaptation Decision}
\label{Adaptation Decision}
In this section, we present the Adaptation Decision procedure, which plays a crucial role in dynamically identifying the most suitable adaptation actions to mitigate the impact of attacks on the workflow. The procedure is described in Algorithm \ref{alg:Adaptation Action Selection Algorithm}.

\vspace*{-.4cm}
\begin{algorithm}
\caption{Adaptation Action Selection Algorithm}
\label{alg:Adaptation Action Selection Algorithm}
\begin{algorithmic} [1]
\REQUIRE $l_{a_k}$: severity level of the detected attack, $BackupSrc_{t_i}$: the backup service.
\ENSURE  Selecting the suitable Adaptation Actions for the detected attack 
\STATE $attackScore \leftarrow ComputeAttackScore(a_k,t_i,s_j,l_{a_k})$ based on Equation \ref{eq:attack_score}
\IF{ $attackScore > Tenant.AdaptTriggerThresh$}
\IF {Lowest-Cost Strategy}
\STATE $finalAA  \leftarrow $  $MA_{l_{(a_k)}}$  $\cap$  $AA_{t_i}$
\FORALL{Adaptation Action $aa$ in $finalAA$}
\STATE $AC[aa]  \leftarrow $ ComputeAdaptationCost($aa$,$t_i$) based on Equation \ref{eq:AdaptationCost}
\ENDFOR
\STATE Sort $AC$ in Ascending order 
\STATE $Selectedaa  \leftarrow$ $aa$ corresponds to the first item in $AC$
\ENDIF
\IF {Adaptive Strategy}
\STATE $Selectedaa \leftarrow$ Predicted $aa$ based on $\mathcal{M}_{ActionSelection}$ (Algorithm \ref{alg:TrainingActionSelectionAlgorithm})
\ENDIF
\IF{ $Selectedaa \in$ tenant-level adaptation actions}
\STATE TenantAdaptation($Selectedaa$)
\ELSE
\STATE MiddlewareAdaptation($Selectedaa$)
\ENDIF

\ENDIF
\end{algorithmic}
\end{algorithm}
\vspace*{-.4cm}

The Adaptation Action Selection algorithm starts by calculating the attack score (line 1), which measures the impact of the detected attack in the current task, taking into account the severity of the attack as well as the security requirements of the task. This computation is based on Equation \ref{eq:attack_score} described in Section \hyperref[sec:4.3.1 Attack Score]{4.3.1}. If the attack score exceeds the pre-defined Adaptation Trigger Threshold ($Tenant.AdaptTriggerThresh$) (line 2), the algorithm proceeds with the adaptation action selection process (lines 3-13).

Two distinct strategies are employed to select the optimal adaptation action: the Lowest-Cost Strategy (lines 3-10) and the Adaptive Strategy (lines 11-13).

The Lowest-Cost Strategy starts by identifying a set of potential adaptation actions, denoted as $finalAA$. This set is determined by intersecting the mitigation actions suitable for the severity of the detected attack ($MA_{l_{(a_k)}}$) and the feasible adaptation actions for the current task ($AA_{t_i}$). The potential actions are then evaluated based on factors such as price, time, mitigation score, and value, computed using Equation \ref{eq:AdaptationCost} described in Section \hyperref[4.3.2 Adaptation Cost]{4.3.2}. The actions are subsequently sorted based on their computed costs, and the one with the lowest cost is selected as the optimal choice for mitigating the attack. 

Furthermore, to address the uncertain costs associated with each adaptation action in the workflow, we propose an Adaptive Strategy. This strategy uses a trained model described in Section \hyperref[4.3.3 Adaptation Model Trainer]{4.3.3}, utilizing historical system reactions and adaptations to violations. This model predicts the optimal adaptation action by considering the current state of the workflow, the previously accrued violations, their corresponding adaptations, and the overall workflow complexity.
By employing this approach, we effectively account for the dynamic and uncertain costs associated with each action, which cannot be statically determined during the adaptation action selection process. In other words, our approach extends beyond the specific task where the attack is detected, encompassing the entire workflow. Through a holistic view, we incorporate any violations occurring in other tasks that may impact the selection of the most suitable adaptation action for the current task. This assessment enables us to make informed decisions regarding the appropriate adaptation action, taking into account the broader scenario and its implications. 

After the selection of the adaptation action(s), the algorithm determines whether they belong to the tenant-level actions or the middleware-level actions. If the action(s) falls under the tenant-level category (line 14), the Tenant Adaptation module is invoked to initiate the necessary adaptations at the tenant level (line 15). Conversely, if the action(s) are categorized as middleware-level adaptations, the Middleware Adaptation module is called upon to implement the required changes (line 17).

\vspace{12pt}
\noindent\textbf{4.3.1 Attack Score}
\label{sec:4.3.1 Attack Score}

The Attack Score serves as a crucial metric for evaluating the impact of a detected attack on the current task and plays a vital role in guiding the selection of appropriate adaptation actions to effectively mitigate its impact on the workflow. The Attack Score is calculated using the following equation:
\vspace*{-.2cm}
\begin{align}
\begin{split}
\label{eq:attack_score}
AttackScore(a_k,t_i,s_j,l_{a_k}) &= (1-\prod_{obj \in \{CIA\}}(1- obj_{t_i}\cdot obj_{a_k})) \\
&\quad \cdot AFR_{s_j}(AT_{a_k})\cdot l_{a_k}
\end{split}
\end{align}

In this equation, $obj_{t_i}$ represents the security requirement of the current task $t_i$ in the workflow for each security objective (CIA, including confidentiality, integrity, and availability). $obj_{a_k}$ denotes the security impact of the detected attack $a_k$ on each security objective. $AFR_{s_j}(AT_{a_k})$ represents the Attack Frequency Rate (AFR) in the cloud service $s_j$ for the type of detected attack $AT_{a_k}$. Lastly, $l_{a_k}$ corresponds to the severity of the detected attack $a_k$ learned by the Algorithm \ref{alg:attack_severity_learning}. 

\vspace{12pt}
\noindent\textbf{4.3.2 Adaptation Cost}
\label{4.3.2 Adaptation Cost}

We define the Adaptation Cost as a metric to evaluate and score potential adaptation actions, with the goal of mitigating the detected attack. It plays a critical role in the Adaptation Decision Engine, enabling evaluation and comparison of the available actions based on the tenant's preferences. The Adaptation Cost takes into account various parameters, such as price, time, mitigation score, and value. It calculates these parameters for each adaptation action and computes the final adaptation cost, considering the weights assigned by the tenant. The Adaptation Cost is calculated using the following equation:
\begin{align}
\begin{split}
AdaptationCost(aa) &= W_{Price} \cdot P_{normalized}(aa) + W_{Time} \cdot T_{normalized}(aa)- \\
&\quad W_{Security} \cdot MS_{normalized}(aa) - W_{Value} \cdot V_{normalized}(aa)
\end{split}
\label{eq:AdaptationCost}
\end{align}
In Equation \ref{eq:AdaptationCost}, the terms $W_{Price}$, $W_{Time}$, $W_{Security}$, and $W_{Value}$ represent the weights assigned by the tenant to price, time, security, and value, respectively, for the given workflow. On the other hand, $P_{normalized}(aa)$, $T_{normalized}(aa)$, $MS_{normalized}(aa)$, and $V_{normalized}(aa)$ denote the normalized price, time, mitigation score, and value, respectively, for the adaptation action $aa$. The time, price, and value are directly assigned based on the adaptation type (refer to Table \ref{table:Adaptation Types}), while the mitigation score is calculated using Equation \ref{Equation:MitigationScore}. This equation considers the security requirements of task $t_i$ (represented by $obj_{t_i}$), the impact of the detected attack $a_k$ on the CIA aspects (represented by $obj_{a_k}$), and the mitigation impact of the adaptation action on each aspect (represented by $obj_{MI_{aa}}$).
\begin{align}
MitigationScore(aa,t_i,a_k) = \sum_{obj \in {C,I,A}} (1 - obj_{t_i} \cdot obj_{a_k}) \cdot obj_{MI_{aa}}
\label{Equation:MitigationScore}
\end{align}

\noindent\textbf{4.3.3 Adaptation Model Trainer}
\label{4.3.3 Adaptation Model Trainer}

In this section, we present our approach for selecting the best adaptation actions using Reinforcement Learning (RL) \cite{kaelbling1996reinforcement}. This approach aims to address the uncertain costs associated with each action, which may impact the entire workflow. These uncertain overhead costs cannot be determined statically at the time of action selection, and it becomes necessary to learn their patterns for each specific workflow. So, we use RL to take a holistic view of the workflow and make informed decisions, considering the unpredictable impact of each adaptation action on other tasks within the workflow. This is particularly important due to the presence of data and control dependencies between tasks in the workflow, as well as the violations that have occurred up to the current state of the workflow and the corresponding adaptations made.

RL is a machine learning approach that deals with decision-making in dynamic environments. We utilize RL within our approach, employing a Markov Decision Process (MDP) to model the decision-making problem. MDP provides a formal framework for representing and solving such problems by defining states, actions, transition probabilities, and rewards. To find optimal policies by learning from previous decision-making experiences, we employ Q-learning \cite{watkins1992q}, a model-free reinforcement learning algorithm.

In the following, we will describe the key elements of the Q-learning problem and present the algorithm for selecting the best adaptation action.

\vspace{8pt}
\noindent\textbf{Markov Decision Process:} A Markov decision process $MDP$ is defined as a 4-tuple $MDP$=($State$, $Action$, $Probability$, $Reward$), with the following definitions:

\begin{itemize}
\item \textbf{$State$}: Represents the set of all possible states. Each state is defined by a 2-tuple $st=(St_T, St_W)$, where $St_T$ denotes the current state of task $t_i$, including the detected attack, and its severity, $St_W$ signifies the present state of the workflow capturing information about the previously occurred violation, their respective adaptations, as well as the time, price, mitigation score, and value of the workflow up to the current point.
\item \textbf{$Action$}: Denotes the set of available actions at a given state. The action set $A(st)$ represents the collection of actions ($a$) that can be taken at state $st$, expressed as $A(st) \subseteq MA_{l_{(a_k)}} \cap AA_{t_i}$ (same as line 4 in Algorithm \ref{alg:Adaptation Action Selection Algorithm}).
\item \textbf{$Probability$}: Describes the probability of transitioning from one state to another when performing a particular action. It is represented by the probability distribution $P(st'|st, a)$.
\item \textbf{$Reward$}: Represents the measure of adaptation action selection efficiency. If action $a$ is selected, the reward function is defined as:
\end{itemize}
\vspace*{-.5cm}
\begin{align}
R(st) = \sum W_i \frac{att_i - att_i^{\min}}{att_i^{\max} - att_i^{\min}}
\label{eq:reward}
\end{align}

In Equation (\ref{eq:reward}), $att_i$ represents the observed values for price, time, value, and mitigation score for the entire workflow, while $att_i^{\max}$ and $att_i^{\min}$ represent the maximum and minimum values of $att_i$ across all adaptation actions. $W_i$ is the weighting factor of $att_i$, where $W_i$ is positive for mitigation score and value and negative for price and time.

The mean Q-value of action $a$ on state $st$ following policy $\pi$ is denoted as $Q_{\pi}(st, a)$. The optimal Q-value function is defined as:
\vspace*{-.2cm}
\begin{align}
Q(st,a) = \sum_{st'} \gamma(st'|st,a) \left[R(st' | st, a) + \gamma \max_{a'} Q(st',a')\right]
\end{align}

Here, $\gamma$ represents the discount factor, $R(st'|st, a)$ is the reward received when transitioning from state $st$ to $st'$ by performing action $a$, and $\max_{a'} Q^*(st',a')$ calculates the maximum Q-value for the next state $st'$. This optimal value function is nested within the Bellman optimality equation.

\vspace*{-.4cm}
\begin{algorithm}
\caption{Action Selection Trainer Algorithm}
\label{alg:TrainingActionSelectionAlgorithm}
\begin{algorithmic} [1]
\REQUIRE $st$: The current state
\ENSURE  Trained Model for Selecting the proper Adaptation Action
\FOR{each episode}
\STATE $st \gets st_0$ 
\FOR{$st \notin St_r $}
\STATE Choose $a \in A(st)$ based on $\epsilon$-greedy policy
\STATE Perform $a$, observe reward $r$ and new state $s'$
\STATE $Q(st,a) \gets Q(st,a)+ \alpha \left[r + \gamma \max_{a'} Q(st',a')- Q(st,a)\right]$
\STATE $st \gets s'$
\ENDFOR
\ENDFOR
\end{algorithmic}
\end{algorithm}

 \section{Evaluation}
 \label{sec:Evaluation}
We implemented \textit{SecFlow}\footnote{Our code is available at \textit{https://github.com/nafisesoezy/SecFlow}} by extending the jBPM (Java Business Process Management) \cite{jBPM} engine and integrating it with the Cloudsim Plus \cite{cloudsimplus} simulation tool. jBPM offers a pluggable architecture that allows for easy replacement of different module implementations. Additionally, the integration of the simulation framework Cloudsim Plus has allowed us to accurately model the complexities of a multi-cloud environment. 

\subsection{Experimental Setting}
To evaluate our approach,  we utilized three distinct categories of process models: Small (3-10 tasks), Medium (10-50 tasks), and Large (50-100 tasks). Our scenario assumed the availability of 5 cloud providers, each offering 3 different services for the service tasks. The specifications of these services fell within the following ranges: Response time [1, 50], Cost [0.1, 10], and confidentiality, integrity, and availability [0, 1]. The response times are selected randomly such that the fastest service is roughly three times faster than the slowest one, and accordingly, it is roughly three times more expensive.

Table \ref{table:Adaptation Types} provides an overview of the relative properties associated with each adaptation type, where $T$, $P$, and $V$ are the original task's response time, price, and value. 
Additionally, $MI$ denotes the mitigation impact of each action on CIA.
\begin{table*}[h!]
\fontsize{8pt}{10pt}\selectfont
\caption{{The Properties of Different Adaptation Types} }
\label{table:Adaptation Types}
\def\arraystretch{1}
\ignorespaces 
\vspace*{-.4cm}
\begin{tabulary}{\linewidth}{ 
p{2.5cm} p{2.8cm} p{2cm} p{2.3cm} p{2cm}}

 \textbf{AdaptType} &
 \textbf{T}  &
\textbf{P}  & 
\textbf{V}  &
\textbf{MI(C,I,A)}
\\  \hline  \hline

\textbf{Insert} &
$T_{newTask}$ & $P_{newTask}$ & $V_{newTask}$ &  $(0.7,0.9,0.9)$
\\  \hline

\textbf{Switch} &
$T_{Switch}$ & $P$  &  $V_{Switch}$ & $(0.7,0.6,0.8)$
\\  \hline

\textbf{Skip} &
$0$  & $0$ &  $0$ & $(0.5,0.4,0.6)$ 
\\  \hline

\textbf{Rework} &
$T_{BackupSrc}$  & $P_{BackupSrc}$ & $V$ & $(0.5,0.9,0.7)$
\\  \hline

\textbf{Redundancy} &
 $Max(T_{BackupSrc},T)$ & $P+P_{BackupSrc}$ & $V+V_{Redundancy}$ & $(0.5,0.8,0.9)$
\\  \hline

\textbf{Reconfiguration} &
$T+T_{reconfig}$ & $P+P_{reconfig}$ & $V+V_{Reconfig}$ & $(0.6,0.7,0.5)$
\\  \hline

\bottomrule 
\end{tabulary}\par 
\vspace*{-.4cm}

\end{table*}

In this paper, we specifically focus on four prevalent types of attacks in cloud services and networks, namely Denial of Service (DoS), probe attacks, Remote-to-Local (R2L), and User-to-Root (U2R). The specifications of these attacks are provided in Table \ref{table:Attack-Specification}. The table presents the Impact on CIA, which indicates the effect of each attack type on CIA security objectives \cite{yang2022network}. It also includes Mitigation Actions, which specify the adaptation actions that effectively mitigate each attack type, classified by attack severity levels (Low, Medium, High).

\begin{table*}[h!]
\fontsize{8pt}{10pt}\selectfont
\caption{{Attack Specifications} }
\label{table:Attack-Specification}
\def\arraystretch{1}
\ignorespaces 
\vspace*{-.4cm}
\begin{tabulary}{\linewidth}{p{1cm}p{2cm}p{2.2cm}p{3cm}p{3.5cm}}
 \textbf{} & \textbf{Impact on} &   \multicolumn{2}{c}{\textbf{Mitigation Actions}} & \textbf{}\\ 
  \textbf{AT} &
 \textbf{ (C,I,A)} &

\textbf{(Low,} &
\textbf{Medium,}&
\textbf{High)} 
\\  \hline  \hline
\textbf{DoS} &
(0.56,0.56,0.56)	&Switch, Rework	&Insert, Rework 	&Insert, Rework, Redundancy, ReConfiguration
\\  \hline

\textbf{Probe} &
(0.22,0.22,0)&	Skip&	 Skip, ReConfiguration& 	Skip, ReConfiguration
\\  \hline

 \textbf{U2R} &
(0.56,0.22,0.22)	&Insert, Rework 	&  Insert, Rework	& Insert, Rework, Redundancy, ReConfiguration
\\  \hline

 \textbf{R2L} &
0.56,0.56,0.22)&	Rework&	Insert, Rework &	 Insert, Rework, ReConfiguration
\\  \hline

\bottomrule 
\end{tabulary}\par 
\vspace*{-.4cm}

\end{table*}
\subsection{Main Results}
In this section, we present the main results of our experiment, focusing on the evaluation of the detection module and the subsequent discussion on the adaptation process.

\noindent \textbf{5.2.1 Detection Method Evaluation} 

\begin{figure*}[htbp]
  \includegraphics[width=\textwidth]{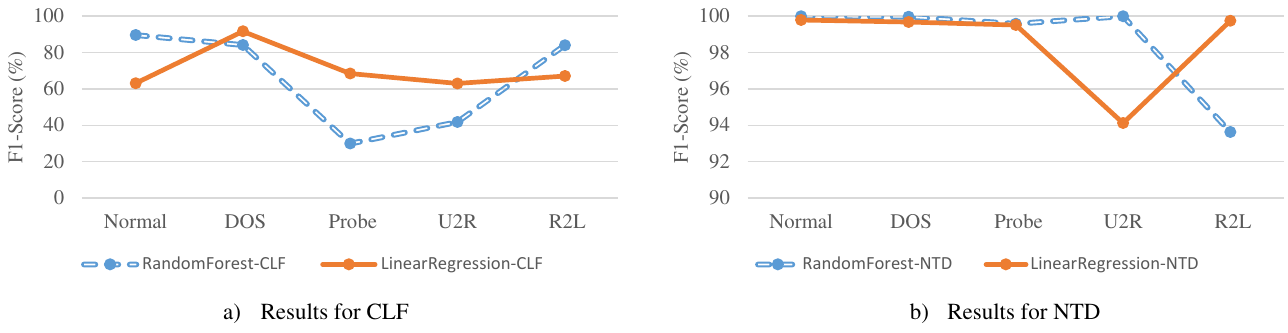}
  \caption{F1-Score Performance of Detection Method for Various Attack Types on CLF and NTD Datasets}
  \captionsetup{skip=0pt}
  \label{fig:F1-Score}
\end{figure*}

We begin by evaluating the performance of our detection method (refer to Algorithm \ref{alg:TraininAttackDetectiongAlgorithm}) using two datasets \footnote{https://github.com/tamaratataru/Bachelors\_Project}: network traffic data (NTD) and cloud log files (CLF). Given the absence of comprehensive existing log files for executing workflows within cloud services, we use synthetic data as follows. To construct the NTD dataset, we simulate various attacks on network traffic data using the KDD dataset \cite{cup1999http} and subsequently integrate it with workflow tasks and cloud service specifications. In a similar way, the CLF dataset is created by simulating attacks within cloud services from different providers, thereby capturing CPU, Bandwidth, and RAM utilization data. Subsequently, the service model trainer module utilizes the Random Forest and Linear Regression algorithms independently to train a model capable of detecting attacks. To evaluate the effectiveness of our detection procedure using these two algorithms, we employ a set of metrics including F1-score, Accuracy, and False Alarm Rate (FAR). Figure \ref{fig:F1-Score} and Table \ref{tab:resultsMl} present the performance comparison between the Random Forest and Linear Regression algorithms for both NTD and CLF datasets. The evaluation demonstrates that the choice between the two methods depends on the specific type of attack being considered. Furthermore, in terms of accuracy, the Random Forest model consistently outperforms Linear Regression across both datasets.

\vspace*{-.4cm}
\begin{table}[h]
\centering

\caption{Detection Accuracy and False Alarm Rate (FAR) for Various Attack Types}
\label{tab:resultsMl}
\begin{tabular}{ p{3.4cm}p{1.6cm}p{1.6cm}p{1.6cm}p{1.6cm}p{1.6cm}}
\hline
DataSet  & Accuracy (\%) & DoSFAR (\%) & ProbeFAR (\%) & R2LFAR (\%) & U2RFAR (\%)\\
\hline
RandomForest-NTD &	99.97&	0.00&	0.03	&0.03&	0.03\\
LinearRegression-NTD  & 99.73	&0.00&	0.01&	0.00	&0.00 \\
RandomForest-CLF  &90.15&	1.56&	3.45&	6.51&	3.45\\
LinearRegression-CLF  &72.07&	5.01&	3.78&	8.53&	5.83\\\hline
\end{tabular}
\end{table}
\vspace*{-.4cm}
\noindent\textbf{5.2.2 Adaptation Method Evaluation} 

In this section, we compare the performance of two adaptation strategies: the Lowest-Cost Strategy and the Adaptive Strategy. We consider uncertain overhead costs introduced by adaptation actions under specific conditions. 

\begin{figure*}[t]
\setlength{\belowcaptionskip}{-12pt}
  \includegraphics[width=\textwidth]{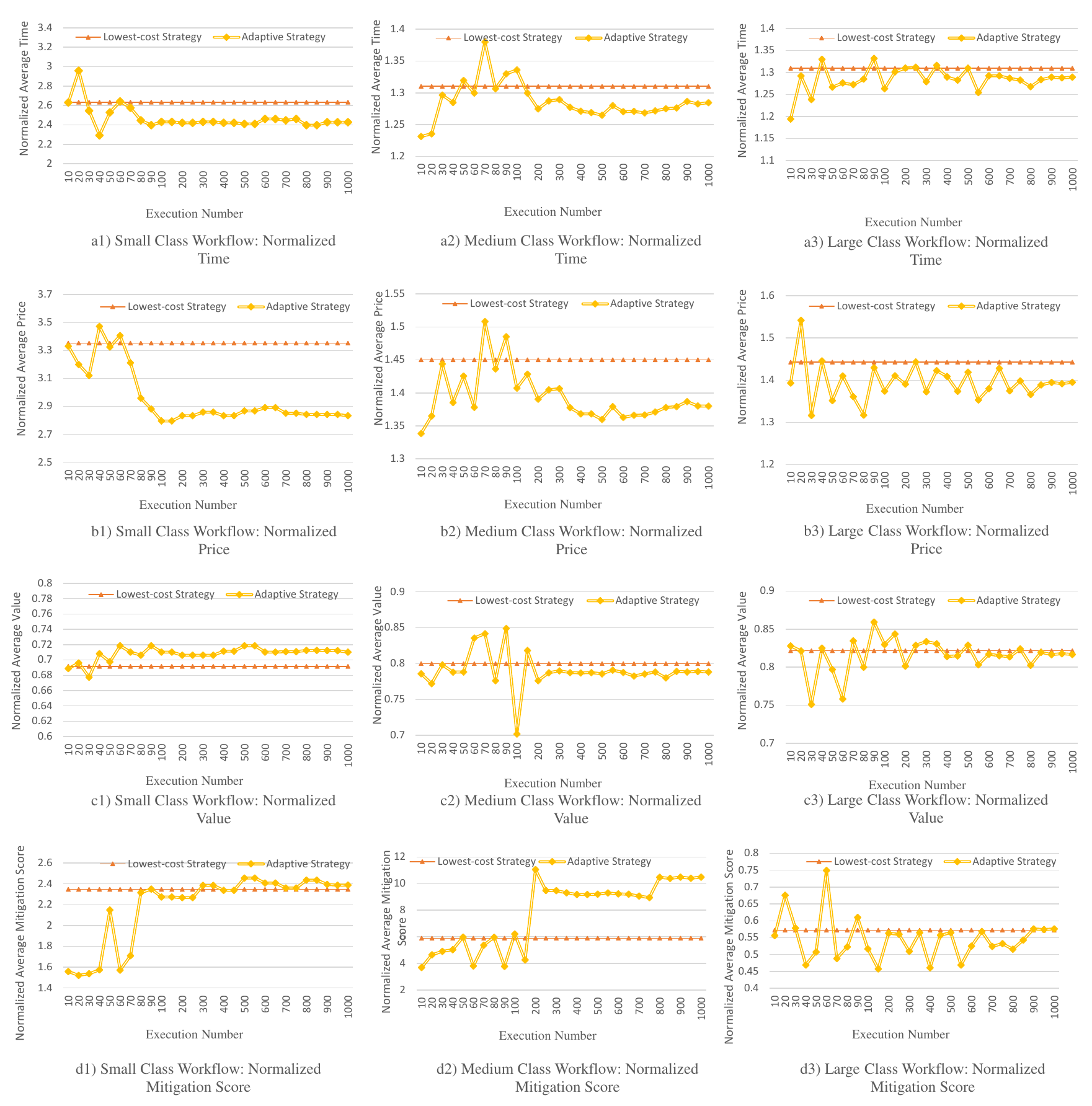}
  \caption{Lowest-Cost vs. Adaptive Strategy Considering Uncertain Overhead Costs}
  \label{fig:finalResult}
\end{figure*}

For the Lowest-Cost Strategy, we calculate the average price, time, and value across 1000 executions of three process categories (small, medium, and large) at an attack rate of 0.3. On the other hand, the Adaptive Strategy is evaluated by calculating the average price, time, and value across every 100 executions over 1000 execution rounds of the three process categories at an attack rate of 0.3.


The results, presented in Figure \ref{fig:finalResult}, show superiority of the Adaptive Strategy for the majority of cases. By intelligently learning the conditions that lead to uncertain costs from adaptation actions in the workflow, it selects the most suitable adaptation action while considering these uncertainties. This selection effectively minimizes the overall execution time and price, while simultaneously maximizing the value and mitigation score. We also observe that finding the optimal set of adaptation actions by the Adaptive Strategy takes longer for the large class workflow compared to the medium and small class workflows due to the larger solution space involved.

\section{Conclusion}
\label{sec:Conclusion}
In this paper, we have addressed critical research gaps in monitoring, detecting, and responding to security violations in cloud-based workflow execution. Our approach focuses on monitoring and detecting security violations, specifically targeting cloud services and network violations. We have presented two strategies for selecting the best action to minimize the impact of such violations. The first strategy selects the most cost-effective adaptation action, while the second leverages adaptive learning from past reactions.

%

To conclude, this paper has established an approach for detecting and subsequently adapting workflows in response to security violations using the introduced strategies. Our approach, implemented as an extension of JBPM and Cloudsim Plus, demonstrated its ability to monitor, detect, and adapt to security violations through simulation results.

In future work, we plan to extend our research to address other potential adversaries, such as tenants and their users, and provide security measures against these attackers. This will further enhance the robustness and effectiveness of our proposed approach in ensuring secure cloud-based workflow execution.

\printbibliography

@inproceedings{SecFlow,
    author={Nafiseh Soveizi et al.},
    title = {SecFlow: Adaptive Security-Aware Workflow Management System in Multi-Cloud Environment},
    year = {2023},
    booktitle={International Conference on Enterprise Design, Operations, and Computing},
    year={2023},
    organization={Springer}
    
    %preprint = {4999533},
    %archivePrefix = {arXiv},
    %url ={http://arxiv.org/abs/2307.05137},
    %primaryClass = {cs.CR} 
}

@article{kaelbling1996reinforcement,
title={Reinforcement learning: A survey},
author={Kaelbling et al.},
journal={J. Artif. Intell. Res.},
volume={4},
pages={237--285},
year={1996}
}

@article{watkins1992q,
  title={Q-learning},
  author={Watkins, Christopher JCH and Dayan, Peter},
  journal={Machine learning},
  volume={8},
  pages={279--292},
  year={1992},
  publisher={Springer}
}

@article{cup1999http,
  title={http://kdd. ics. uci. edu/databases/kddcup99/kddcup99. html},
  author={Cup, KDD},
  journal={The UCI KDD Archive},
  year={1999}
}

@article{yang2022network,
  title={Network security situation assessment with network attack behavior classification},
  author={Yang, Hongyu and Zhang, Zixin and Xie, Lixia and Zhang, Liang},
  journal={International Journal of Intelligent Systems},
  volume={37},
  number={10},
  pages={6909--6927},
  year={2022},
  publisher={Wiley Online Library}
}

@inproceedings{nolle2018binet, title={BINet: multivariate business process anomaly detection using deep learning}, author={Nolle et al.}, booktitle={International Conference on Business Process Management}, pages={271--287}, year={2018}, organization={Springer}
}

@article{soveizi2023security,
  title={Security and privacy concerns in cloud-based scientific and business workflows: A systematic review},
  author={Soveizi, Nafiseh and Turkmen, Fatih and Karastoyanova, Dimka},
  journal={Future Generation Computer Systems},
  year={2023},
  publisher={Elsevier}
}

@article{chen2015towards,
  title={Towards energy-efficient scheduling for real-time tasks under uncertain cloud computing environment},
  author={Chen, Huangke and Zhu, Xiaomin and Guo, Hui and Zhu, Jianghan and Qin, Xiao and Wu, Jianhong},
  journal={Journal of Systems and Software},
  volume={99},
  pages={20--35},
  year={2015},
  publisher={Elsevier}
}

@inproceedings{chen2016uncertainty,
  title={Uncertainty-aware real-time workflow scheduling in the cloud},
  author={Chen, Huangke and Zhu, Xiaomin and Qiu, Dishan and Liu, Ling},
  booktitle={2016 IEEE Cloud Conference},
  pages={577--584},
  %year={2016},
  organization={IEEE}
}

@article{nolle2018analyzing,
  title={Analyzing business process anomalies using autoencoders},
  author={Nolle, Timo and Luettgen, Stefan and Seeliger, Alexander and M{\"u}hlh{\"a}user, Max},
  journal={Machine Learning},
  volume={107},
  pages={1875--1893},
  year={2018},
  publisher={Springer}
}

@misc{jBPM,
    title = {jBPM: Business Process Management Suite},
    howpublished = {\url{https://www.jbpm.org/}}
}

@misc{cloudsimplus,
title={{CloudSim Plus}},
author={{CloudSim Plus Contributors}},
howpublished={{GitHub repository}},
note={\url{https://github.com/manoelcampos/cloudsim-plus}}
}

@inproceedings{Maguluri2012,
author = {Maguluri, Siva Theja and others},
booktitle = {Proceedings IEEE Infocom},
pages = {702--710},
title = {{Stochastic models of load balancing and scheduling in cloud computing clusters}},
year = {2012}
}

@article{HosseiniShirvani2020,
abstract = {Cloud computing became an inevitable information technology industry. Despite its several plus points such as economy of scale and rapid elasticity, it suffers from vendor lock-in, resource limitation and cybersecurity attacks in which it leads business discontinuity or even business failure. Multi-cloud, on the other hand, can be trustable paradigm to obviate obstacles such as aforesaid unpleasant features of a single cloud. One of the biggest challenges is to know which cloud is commensurate with user's business process with regards to security objectives. To this end, the new method is presented to quantify the amount of cloud security risk (CSR) in regards to user's business process. Therefore, in this paper, the web service composition problem is formulated to bi-objective optimisation problem with service cost and multi-cloud risk viewpoints in ever-increasing multi-cloud environment (MCE) in which each provider has its variable pricing policy and different security level. It is obviously an NP-Hard problem. To solve the combinatorial problem, we develop a bi-objective time-varying particle swarm optimisation (BOTV-PSO) algorithm. The parameters are tuned based on elapsed time so a good balance between exploration and exploitation is achieved. To illustrate the effectiveness of proposed algorithm, we defined several scenarios and compared the performance of proposed algorithm with multi-objective GA-based (MOGA) optimiser, a single objective genetic algorithm (SOGA) that only optimises cost function and neglects CSR, and multi-objective simulated annealing algorithm (MOSA). The experimental results showed the superiority of proposed BOTV-PSO against other approaches in terms of convergence, diversity, fitness, performance, and even scalability.},
author = {{Hosseini Shirvani}, Mirsaeid},
%doi = {10.1080/0952813X.2020.1725652},
file = {:C\:/Users/MWP-/Documents/nafas/nafiseh phd/survey/new/search results/web of science/bp/scheduling/Bi-objective web service composition problem in multi-cloud environment-a bi-objective time-varying particle swarm optimisation algorithm.pdf:pdf},
%issn = {13623079},
journal = {Journal of Experimental and Theoretical Artificial Intelligence},
keywords = {Multi-objective optimisation,multi-cloud market,web service composition},
%number = {00},
pages = {1--24},
publisher = {Taylor & Francis},
title = {{Bi-objective web service composition problem in multi-cloud environment: a bi-objective time-varying particle swarm optimisation algorithm}},
%url = {https://doi.org/10.1080/0952813X.2020.1725652},
%volume = {00},
year = {2020}
}

@article{Ahmad2021,
author = {Ahmad, Zulfiqar and Nazir, Babar and Umer, Asif},
%doi = {10.1002/dac.4649},
file = {:C\:/Users/MWP-/Documents/nafas/nafiseh phd/survey/new/writing survey/new related paper/new should check paper/latestpublishpaperzulfiqar.pdf:pdf},
%issn = {10991131},
journal = {International Journal of Communication Systems},
keywords = {CyberShake,Montage,admission,engine,fault tolerant,mapper,scheduler,scheduling,scientific workflows},
number = {1},
title = {{A fault-tolerant workflow management system with Quality-of-Service-aware scheduling for scientific workflows in cloud computing}},
volume = {34},
year = {2021}
}

@inproceedings{Varshney2019,
author = {Varshney, Shefali and others},
booktitle = {International Conference on Advances in Computing and Data Sciences},
pages = {711--723},
title = {{QoS Based Resource Provisioning in Cloud Computing Environment: A Technical Survey}},
year = {2019}
}

@article{Wang2020b,
author = {Wang et al.},
journal = {IET Information Security},
volume = {14},
number = {2},
pages = {157--165},
title = {{Protecting scientific workflows in clouds with an intrusion tolerant system}},
year = {2020}
}

@article{Abazari2019,
abstract = {Cloud computing is emerging with growing popularity in workflow scheduling, especially for scientific workflow. Deploying data-intensive workflows in the cloud brings new factors to be considered during specification and scheduling. Failure to establish intermediate data security may cause information leakage or data alteration in the cloud environment. Existing scheduling algorithms for the cloud disregard the interaction among tasks and its effects on application security requirements. To address this issue, we design a new systematic method that considers both tasks security demands and interactions in secure tasks placement in the cloud. In order to respect security and performance, we formulate a model for task scheduling and propose a heuristic algorithm which is based on task's completion time and security requirements. In addition, we present a new attack response approach to reduce certain security threats in the cloud. To do so, we introduce task security sensitivity measurement to quantify tasks security requirements. We conduct extensive experiments to quantitatively evaluate the performance of our approach, using WorkflowSim, a well-known cloud simulation tool. Experimental results based on real-world workflows show that compared with existing algorithms, our proposed solution can improved the overall system security in terms of quality of security and security risk under a wide range of workload characteristics. Additionally, our results demonstrate that the proposed attack response algorithm can effectively reduce cloud environment threats.},
author = {Abazari, Farzaneh and Analoui, Morteza and Takabi, Hassan and Fu, Song},
%doi = {10.1016/j.simpat.2018.10.004},
file = {:C\:/Users/MWP-/Documents/nafas/nafiseh phd/survey/new/search results/web of science/scientific/scheduling/read/MOWS-Multi-objective workflow scheduling in cloud computing based on heuristic algorithm.pdf:pdf},
%issn = {1569190X},
journal = {Simulation Modelling Practice and Theory},
keywords = {Attack response,Cloud computing security,Scientific workflows,Secure task scheduling},
number = {October 2018},
pages = {119--132},
publisher = {Elsevier},
title = {{MOWS: Multi-objective workflow scheduling in cloud computing based on heuristic algorithm}},
%url = {https://doi.org/10.1016/j.simpat.2018.10.004},
volume = {93},
year = {2019}
}

@article{Alaei2021b,
abstract = {With the increasing popularity and acceptance of cloud computing, it is being applied in services like executing large-scale applications, where cloud environment is selected by the scientific associations to easily execute the computation intensive workflows. However, cloud computing can have higher failure rates due to the larger number of servers and components filled with the intensive workloads. These failures may lead to the unavailability of virtual machines (VMs) for computation. Hence, this issue of fault occurrences can be tolerated by adopting an effective and efficient fault tolerant strategy. The goal of our research in this paper is to develop an adaptive fault detector strategy based on Improved Differential Evolution (IDE) algorithm in cloud computing that can minimize the energy consumption, the makespan, the total cost and, at the same time, tolerate up faults when scheduling scientific workflows. This proposed work applies an adaptive network-based fuzzy inference system (ANFIS) prediction model to proactively control resource load fluctuation that increases the failure prediction accuracy before fault/failure occurrence. In addition, it applies a reactive fault tolerance technique for when a processor fails and the scheduler must allocate a new VM to execute the workflow tasks. The experimental results show that compared with existing techniques, the proposed approach significantly improves the overall scheduling performance, achieves a higher degree of fault tolerance with high HyperVolume (HV) compared with the ICFWS, IDE, and ACO algorithms, minimizes the makespan, the energy consumption and task fault ratio, and reduces the total cost.},
author = {Alaei, Mani and others},
%doi = {10.1016/j.asoc.2020.106895},
file = {:C\:/Users/MWP-/Documents/nafas/nafiseh phd/chapter plan/chapter3/Architecture/attack pic/detecting attack in scientific workflow/Alaei-2021-An-adaptive-fault-detector-strategy.pdf:pdf},
%issn = {15684946},
journal = {Applied Soft Computing},
keywords = {ANIFS,Cloud computing,Fault tolerance,Migration,Workflow scheduling},
pages = {106895},
publisher = {Elsevier B.V.},
title = {{An adaptive fault detector strategy for scientific workflow scheduling based on improved differential evolution algorithm in cloud}},
%url = {https://doi.org/10.1016/j.asoc.2020.106895},
volume = {99},
year = {2021}
}

@article{Wang2021,
author = {Wang, Yawen and Guo, Yunfei and Wang, Wenbo and Liang, Hao and Huo, Shumin},
%doi = {10.1016/j.future.2020.08.004},
file = {:C\:/Users/MWP-/Documents/nafas/nafiseh phd/survey/new/writing survey/new related paper/INHIBITOR- An intrusion tolerant scheduling algorithm in cloud-based.pdf:pdf},
%issn = {0167739X},
journal = {Future Generation Computer Systems},
keywords = {Intrusion tolerance,Scientific workflow security,Task replication,Workflow scheduling},
pages = {272--284},
publisher = {Elsevier B.V.},
title = {{INHIBITOR: An intrusion tolerant scheduling algorithm in cloud-based scientific workflow system}},
%url = {https://doi.org/10.1016/j.future.2020.08.004},
volume = {114},
year = {2021}
}

@article{Wen2020b,
author = {Wen et al.},
journal = {IEEE Transactions on Cloud Computing},
title = {{Dynamically Partitioning Workflow over Federated Clouds for Optimising the Monetary Cost and Handling Run-Time Failures}},
year = {2020},
pages = {1093--1107},
volume = {8},
number = {4},
}
\end{document}